\documentclass{article}
\usepackage{amsfonts}
\usepackage{amssymb}
\usepackage{amsbsy}
\usepackage{amsthm}
\usepackage{amsmath}
\newtheorem{lemma}{Lemma}
\newtheorem{prop}[lemma]{Proposition}
\newtheorem{theorem}[lemma]{Theorem}

\newcommand{\bibtitle}[1]{\textit{#1}}
\newcommand{\bibseries}[1]{\textbf{#1}}
\newcommand{\bibyear}[1]{(#1)}

\newcommand{\ioo}[1]{\ensuremath{\:\left]#1\right[\:}}
\newcommand{\ico}[1]{\ensuremath{\:\left[#1\right[\:}}
\newcommand{\ltag}[1]{\tag*{\hbox to 0pt{\hss\hbox to \displaywidth{(#1)\hss}}}}

\newcommand{\setr}{\ensuremath{\mathbb{R}}}
\newcommand{\setrq}{\ensuremath{\bar{\mathbb{R}}}}
\newcommand{\g}{\ensuremath{\mathfrak{g}}}
\newcommand{\T}{\ensuremath{\mathcal{T}}}
\newcommand{\two}{\ensuremath{I\!\!I}}
\newcommand{\dt}{\ensuremath{\partial_t}}
\newcommand{\tr}[1]{\ensuremath{\mathrm{tr}_{#1}\,}}
\newcommand{\Div}[1]{\ensuremath{\mathrm{div}_{#1}\,}}

\newcommand{\del}{\ensuremath{\nabla}\!}
\newcommand{\jrd}[1]{\ensuremath{\tilde{#1}}}

\begin{document}

%
%

\author{Roger Bieli\\
Max Planck Institute for Gravitational Physics\\
Am M\"uhlenberg 1\\
14476 Golm\\
Germany
}

\title{Coupled quintessence and curvature-assisted acceleration}
\date{}

\maketitle

\begin{abstract}
 Spatially homogeneous models with a scalar field non-minimally coupled to
 the space-time curvature or to the ordinary matter content are analysed with
 respect to late-time asymptotic behaviour, in particular to accelerated
 expansion and isotropization. It is found that a direct coupling to the
 curvature leads to asymptotic de Sitter expansion in arbitrary exponential
 potentials, thus yielding a positive cosmological constant although
 none is apparent in the potential. This holds true regardless of the
 steepness of the potential or the smallness of the coupling constant. For
 matter-coupled scalar fields, the asymptotics are obtained for a large
 class of positive potentials, generalizing the well-known cosmic no-hair
 theorems for minimal coupling. In this case it is observed that the direct
 coupling to matter does not impact the late-time dynamics essentially.
\end{abstract}

%
%

\section{Introduction}

The term \emph{coupled quintessence} was introduced by Amendola \cite{Ame00}
and refers to a non-linear real scalar field within a cosmological model
that is in general non-minimally coupled to the other forms of matter. Such
a field is commonly used as a generalization of a cosmological constant
or ordinary (minimally coupled) quintessence \cite{Cal98} to account for
the observed late-time acceleration of the universe. Phenomenologically,
the main motivation for considering an explicit coupling to matter is
the existence of scaling attractors that exhibit accelerated expansion
\cite{Ame99,Ame00,Ame01,Bil00,Gum05,Zim01}, a feature which is closely
related to the \emph{cosmic coincidence} problem \cite{Ste97}. From the
mathematical point of view, such couplings are interesting because they
arise from the conformal transformation of scalar-tensor theories from the
Jordan frame, where the scalar field is minimally coupled to matter, to the
Einstein frame, in which the scalar field is minimally coupled to gravity,
though at the expense of an explicit coupling to matter. Thus, knowledge
on the dynamics in the Einstein frame can lead to an understanding of the
behaviour of a large class of conformally related scalar-tensor or higher
order gravity theories, prominent examples thereof may include Brans--Dicke
theory \cite{Bra61} and $f(R)$-gravity \cite{Bar88}.

A specific scalar-tensor theory belonging to the above-mentioned class that
gained a lot of attention is obtained by directly coupling the quintessence
field to the scalar curvature of the space-time. Such a coupling was
originally introduced for the energy-momentum tensor of a self-interacting
scalar particle to become renormalizable \cite{Cal70}. Models of this
kind were studied with respect to different inflationary scenarios
\cite{Ame90,Far96,Far01,Fut89a,Fut89b,Saa01} and were engaged to address
the \emph{cosmological constant} problem \cite{Mad88,Mat04} or to violate
the null energy condition in order to obtain a barotropic index $w < -1$
for the corresponding perfect fluid \cite{Saa01}. Noteworthy too is the
existence of stable scaling solutions in potentials other than exponential
as found by Uzan \cite{Uza99} for the inverse power-law case. Recently,
Tsujikawa \cite{Tsu00} pointed out a particularly interesting consequence
of a non-minimal coupling between scalar field and curvature, namely
the possibility to obtain inflationary solutions in an exponential
potential that is too steep to sustain inflation in the minimally coupled
case. Following terminology for a similar effect in multi-field inflation
models \cite{Lid98}, this behaviour was called \emph{curvature-assisted
inflation}.

Quintessence models, minimally coupled or not, require the specification
of a potential, in which the scalar field evolves. In work pioneered by
Starobinsky \cite{Sta98} it was shown that, in principle, it is possible to
reconstruct the quintessence potential from a given expansion history of
the universe.  While, phenomenologically, this certainly is a comfortable
situation, it also means that basically any compatible dynamics can be
obtained by just choosing an appropriate potential. As a consequence of
this ``scalar Synge trick'', instead of regarding the potential as a free
function, it seems physically reasonable to require some motivation for
it. Exponential potentials can provide this easily, as they naturally arise
from Kaluza-Klein type reductions of higher dimensional theories or from
conformal rescalings, {\it e.g.} of a string frame cosmological constant. The
existence of scaling attractors \cite{Bil98,Cop98} and power-law inflationary
solutions \cite{Hal87} also rendered them attractive for cosmology.

The aim of the present paper is twofold. First, the \emph{cosmic no-hair}
theorems of Wald \cite{Wal83} for a positive cosmological constant,
of Kitada and Maeda \cite{Kit92,Kit93} for quintessence in an exponential
potential and of Rendall \cite{Ren04,Ren05,Ren06} for non-linear scalar
fields in more generic potentials shall be generalized to the case of
coupled quintessence. Those theorems concern initially expanding global
solutions of Bianchi type I to VIII in the presence of ordinary matter and
establish late-time acceleration and isotropization. Second, these results
shall be applied to a scalar field in an exponential potential which is
non-minimally coupled to the space-time curvature to demonstrate that
an arbitrarily small positive coupling constant allows for an asymptotic
de Sitter expansion, no matter how steep the potential actually is.

In section \ref{sec:kin} the notation is set up and the equations of motion
for spatially homogeneous models of coupled quintessence are obtained.
Section \ref{sec:exp} states the main assumptions, gives some basic
estimates and derives the asymptotics in the case where the potential energy
density of the field is bounded below by a positive constant, leading
to exponential acceleration and isotropization. Section \ref{sec:int}
in contrast deals with evolutions where the potential energy density is
allowed to approach zero which leads to a more subtle dynamics and, in
general, intermediate acceleration only. Curvature-assisted acceleration
is presented in section \ref{sec:caa} for scalar fields non-minimally
coupled to the scalar curvature of space-time by conformally relating them
to the coupled quintessence models considered beforehand. Finally, section
\ref{sec:con} summarizes the results and concludes with some comments.

%
%

\section{Kinetics of coupled quintessence} \label{sec:kin}

The purpose of this section is to fix notations, to state the coupled
quintessence model explicitly for the spatially homogeneous case and to
obtain a $n+1$-decomposition of the equations of motion with respect to the
hypersurfaces of homogeneity. For this, let $G$ be a Lie group of dimension
$n \geq 3$ with Lie algebra $\g$, $I \subset \setr$ an interval in the
reals $\setr$ with non-empty interior and $\gamma \in C^2(I,T^0_2\g)$
a family of Riemannian metrics on $\g$. Then, if $M:=G \times I$ with the
canonical projections $\pi:M \to G$ and $t:M \to I$, \[ g := \pi^\ast \gamma
- dt \otimes dt \] is a Lorentzian metric on $M$, where the pullback of an
arbitrary family $F \in C(I,T^0_s\g)$ is defined by $(\pi^\ast F)(p,t):=
(\pi^\ast F(t))(p,t) \in \T^0_s M$ for all $p \in G$, $t \in I$. Here, $F(t)
\in T^0_s\g$ is identified with its $C^\infty(G)$-multilinear left-invariant
extension to $\T^0_s G$ and $\T^r_s M$ stands for the set of continuous
sections of the $(r,s)$-tensor bundle $T^r_s M$ over $M$. Having \[k:= -
\frac12 \dot g \in C^1(I,T^0_2\g)\] the second fundamental form $\two$
of a $t$-hypersurface in $M$ is given by \[ \two(X,Y) = - k(t)(T\pi X,T\pi
Y) \dt \ , \qquad X,Y \in \T(G \times \{ t \}).\] A common convention in
cosmology is to denote the \emph{negative} of the mean curvature by $H$,
so $H:=-\tr\gamma k / n$.

Similar to the metric on $M$, the energy-momentum tensor $T$ of the ordinary
matter content is constructed by choosing an energy density $\rho \in
C^1(I)$, a flow covector $j \in C^1(I,T^0_1\g)$ and a symmetric pressure
tensor $S \in C^1(I,T^0_2\g)$ and setting \[T := \pi^\ast S - \pi^\ast j
\otimes dt - dt \otimes \pi^\ast j + \rho\, dt \otimes dt.\] For this indeed
to describe ordinary matter, energy conditions will be imposed later on,
but for the moment it is enough to know the decomposition given above.

Finally, for an interval $J \subset \setr$ with non-empty interior fix coupling
functions $C \in C^1(J)$ and $c \in C(J)$ as well as a potential $V \in
C^1(J)$ for the quintessence field $\phi \in C^2(I,J)$. Then, the coupled
Einstein-scalar~field-matter (ESM) equations on $M$ read
\begin{align}
 Rc_g - \frac12 R_g g & = \del_g \phi \otimes \del_g \phi - \frac12 |\del_g
  \phi|^2_g g - V(\phi) g + C(\phi) T \label{feseq} \\
 \Box_g \phi - V'(\phi) & = - c(\phi) \tr g T \label{fsfeq} \\
 \Div g C(\phi)T & = c(\phi) (\tr g T) \del_g \phi \label{fmteq}
\end{align}
with $Rc$ and $R$ are Ricci tensor and scalar, $\del$ is the Levi-Civita
connection and $\Box$ the covariant wave operator on $(M,g)$. Compositions
with $\pi$ and $t$ are understood implicitly where appropriate.  Note the
particular kind of direct coupling assumed between the scalar field $\phi$
and the matter energy-momentum tensor $T$ in the right hand sides of these
equations. Minimal coupling would correspond to $C=1$ and $c=0$ identically.
While this is, of course, not the most general form of coupling conceivable,
it encompasses the cases studied \emph{e.g.} in \cite{Ame00} or \cite{Gum05}
and is sufficient for the application to curvature-coupled models in
section \ref{sec:caa}.

The $n+1$-decomposition of the ESM equations results in the Hamiltonian
and momentum constraints
\begin{align}
 n(n-1) H^2 & = \dot\phi^2 + 2 V(\phi) + |\sigma|_\gamma^2 - R_\gamma +
 2 C(\phi) \rho \label{hamcon} \\
 \Div\gamma \sigma & = C(\phi)j, \label{momcon}
\end{align}
the evolution equations
\begin{align}
 n(n-1) \dot H & = -n \dot\phi^2 -n|\sigma|_\gamma^2 + R_\gamma -
 C(\phi) (n\rho + \tr\gamma S) \label{hevl} \\
 \dot\sigma & = -2 \sigma\cdot\sigma -(n-2) H\sigma + \hat{Rc}_\gamma -
 C(\phi) \hat{S}, \label{sgmevl}
\end{align}
the scalar field equation
\begin{equation} \label{sfeq}
 \ddot\phi + n H \dot\phi + V'(\phi) = c(\phi) \tr g T
\end{equation}
and the equations of motion for the ordinary matter
\begin{align}
 \left[ C(\phi)\rho \right]\dot{} + \Div\gamma C(\phi)j - C(\phi) \langle
  \sigma,\hat{S} \rangle + H C(\phi)(n \rho + \tr\gamma S) & = - c(\phi)(\tr
  g T) \dot\phi \label{eom} \\
 \left[ C(\phi)j \right]\dot{} + \Div\gamma C(\phi)\hat S + nHC(\phi)j &
  = 0 \label{eom2}
\end{align}
A hat shall denote the trace-free part of a tensor, $\sigma:=k+H\gamma$
is the shear tensor and $\langle\cdot,\cdot\rangle_\gamma$ is the family of
fibre metrics induced by $\gamma$ on the tensor bundle $T^0_2\g$ with the
abbreviation $|\sigma|_\gamma^2 := \langle\sigma,\sigma\rangle_\gamma$, while
$(\sigma \cdot \sigma)(t)(X,Y) := \tr{\gamma(t)} \big[ \sigma(t)(X,\cdot)
\otimes$ $\sigma(t)(\cdot,Y) \big]$ for all $t \in I$ and $X,Y \in \g$.

%
%

\section{Exponential acceleration} \label{sec:exp}

To analyse the asymptotics of global solutions of the coupled quintessence
model stated in section \ref{sec:kin} some assumptions on the Lie group
$G$, the potential $V$, the coupling functions $c$ and $C$ as well as on
the energy-momentum tensor $T$ describing the ordinary matter content
will now be made. Two cases are then considered separately, namely one
where the potential includes a positive cosmological constant that leads
asymptotically to exponential inflation, and a second in the next section
where the potential energy density of the scalar field approaches zero,
which allows, in general, for intermediate late-time acceleration only. The
argument closely follows that of \cite{Ren04}.

The Lie group $G$ is assumed such that every left invariant Riemannian
metric has non-positive scalar curvature. In the three-dimensional simply
connected case, this corresponds to groups of Bianchi type other than IX
\cite{Wal83}. It follows now from the outline of section \ref{sec:kin}
that $R_\gamma \leq 0$.  The energy-momentum tensor $T$ is supposed to
satisfy the dominant and strong energy condition, which imply
\begin{gather*}
 |j|_\gamma \leq \rho,\ |\tr\gamma S| \leq n \rho \ltag{DEC} \\
 \rho + \tr\gamma S \geq 0. \ltag{SEC}
\end{gather*}
Furthermore, it is assumed that the coupling function $C$ is non-negative,
so $C(\phi)T$ fulfills every energy condition $T$ does, and that there is
a constant $C_0$ such that
\begin{align*}
 |c| \leq C_0 C \ltag C
\end{align*}
holds. The potential $V$ shall be positive and satisfy the following, more
technical conditions:
\begin{itemize}
 \item[(P1)] There exists a positive lower semi-continuous minorant $\bar V$
  on the closure $\bar J$ of $J$ in $\setr$, \emph{i.e.} $\bar V : \bar J \to
  \setr$ is lower semi-continuous, $\bar V >0$ and $V(x) \geq \bar V(x)$
  for all $x \in J$
 \item[(P2)] The derivative $V'$ is bounded on any subset of $J$ on which
  $V$ is bounded.
 \item[(P3)] $V'$ extends to a continuous function on the closure of $J$
  in $\setrq:=\setr \cup \{\pm \infty\}$ with values in $\setrq$.
\end{itemize}
(P1) ensures that $V$ is bounded away form zero on bounded subsets of $J$
whereas (P3) yields the existence of limits $\lim_{x \to J_\pm} V'(x)=
V'_\pm \in \setrq$ when $x$ approaches the endpoints $J_\pm$ of $J$
in $\setrq$.  Finally, suppose that the model is initially expanding, so
$t_0=\min I \in \setr$ and $H(t_0)>0$. Lemma \ref{bounds} below collects
some immediate consequences of these assumptions.

\begin{lemma} \label{bounds}
 The following properties hold:
 \begin{enumerate}
  \item $H$ is positive, bounded and monotonically decreasing
  \item $\dot\phi$ is bounded and square-integrable with
   $\|\dot\phi\|_{L^\infty} \leq \sqrt{n(n-1)} \|H\|_{L^\infty}$ and
   $\|\dot\phi\|^2_{L^2} \leq (n-1) \|H\|_{L^\infty}$
  \item $\ddot\phi$, $R_\gamma$, $|\sigma|^2_\gamma$, $V(\phi)$, $V'(\phi)$,
   $C(\phi)\rho$ and $c(\phi)\tr g T$ are bounded with
   \begin{gather*}
    \|R_\gamma\|_{L^\infty}, \||\sigma|^2_\gamma\|_{L^\infty},
    2\|V(\phi)\|_{L^\infty}, 2\|C(\phi)\rho\|_{L^\infty} \leq n (n-1)
    \|H\|^2_{L^\infty} \\
    \|c(\phi)\tr g T\|_{L^\infty} \leq \frac12n(n-1)(n+1)C_0
    \|H\|_{L^\infty}
   \end{gather*}
 \end{enumerate}
\end{lemma}

\begin{proof}
 Since $V$ is positive so is $H^2$ due the Hamiltonian constraint
 \eqref{hamcon} and so is $H$ itself because it is positive initially. The
 evolution equation \eqref{hevl} and (DEC) render $\dot H$ non-positive
 in the same way, so that $H$ is monotonically decreasing and bounded by
 $0 < H \leq H(t_0)$. In fact, the inequality $\dot\phi^2 \leq - (n-1)
 \dot H$ holds, so integration over $I$ yields $\|\dot\phi\|^2_{L^2} \leq
 (n-1) \|H\|_{L^\infty}$. The $L^\infty$ bounds on $\dot\phi$, $R_\gamma$,
 $|\sigma|^2_\gamma$, $V(\phi)$ and $C(\phi)\rho$ follow form \eqref{hamcon}
 directly, $V'(\phi) \in L^\infty(I)$ is then a consequence of (P2) and the
 estimate for $c(\phi) \tr g T$ is obtained from the inequality $|c(\phi)
 \tr g T|=|c(\phi)| |\tr\gamma S - \rho| \leq C_0 C(\phi) (n+1) \rho$ using
 (C). The scalar field equation \eqref{sfeq} finally gives a $L^\infty$
 bound for $\ddot\phi$.
\end{proof}

Since $V$ is a positive function, the infimum of the potential energy
density of the field $V_0 := \inf\,(V \circ \phi)(I)$ is either positive or
zero. The rest of this section is concerned with the late-time asymptotics
in this first case. It is shown that expansion, isotropization and decay
of matter terms take place exponentially in time. Note that the solution
is assumed to exist globally, so $I$ is not bounded from above.

\begin{prop} \label{mtdecay}
 For any $\delta>0$ the spatial curvature $R_\gamma$, the shear
 $|\sigma|^2_\gamma$, the matter terms $C(\phi)\rho$, $C(\phi)|j|_\gamma$
 and $C(\phi) \tr\gamma S$ as well as the coupling term $c(\phi)
 \tr g T$ decay at least as $e^{-(2-\delta) H_0 t}$, \emph{i.e.}
 \[R_\gamma,\,|\sigma|^2_\gamma,\,C(\phi)\rho,\,C(\phi)|j|_\gamma,\,C(\phi)
 \tr\gamma S,\,c(\phi) \tr g T = \mathcal{O}\left(e^{-(2-\delta) H_0
 t}\right) \quad (t \to \infty),\] where $H_0:=\sqrt{ 2 V_0 / n(n-1)}.$
\end{prop}

\begin{proof}
 (i) From the square-integrability of $\dot\phi$ and the boundedness of
 $\ddot\phi$ (see Lemma \ref{bounds}) it follows that $\dot\phi(t) \to 0$
 as $t \to \infty$. (ii) With $E := 1/2\,\dot\phi^2 + V(\phi) \in C^1(I)$
 the field energy density, define a quantity \cite{Ren04} \[ Z :=
 n(n-1)H^2-2E \in C^1(I). \] Because of \eqref{sfeq} $\dot E = c(\phi)(\tr g
 T)\dot\phi -nH\dot\phi^2$, which implies with (SEC) 
 \begin{align*}
  \dot Z & =-2H \left( Z + (n-1) |\sigma|^2_\gamma + C(\phi)\left[ (n-2)
   \rho + \tr\gamma S \right] \right) -2c(\phi)(\tr g T)\dot\phi \\
  & \leq -2HZ -2 c(\phi)(\tr g T)\dot\phi.
 \end{align*}
 (iii) Fix a $0<\delta<1$, then \[ \delta H C(\phi)\rho + 2 c(\phi)(\tr g
 T)\dot\phi \geq \left[ \delta H -2(n+1) C_0 |\dot\phi| \right] C(\phi)\rho
 \] holds and with $-2HZ \leq -(2-\delta)HZ - \delta H C(\phi)\rho$
 and $H\geq H_0$ by \eqref{hamcon} one obtains \[ Z \leq - (2-\delta)
 H_0 Z \quad {\rm eventually}\] from (i). Integrating this yields \[ Z(t)
 = \mathcal{O}\left(e^{-(2-\delta) H_0 t}\right) \qquad (t\to\infty).\]
 Noting again that $Z=|\sigma|^2_\gamma-R_\gamma+C(\phi)\rho$ and using
 (DEC) and (C) now establishes the claims.
\end{proof} 

With this information at hand, it is possible to obtain late-time limits for
the mean curvature $-H$, the potential energy density of the field $V(\phi)$,
its derivative, and for the derivatives of the field $\phi$ itself.

\begin{prop} \label{flddecay}
 As $t$ goes to infinity, the following limits are attained:
 \begin{enumerate}
  \item $H(t) \to H_\infty$, with $H_\infty \geq H_0 > 0$
  \item $(V\circ\phi)(t) \to V_\infty := \frac12 n(n-1)H^2_\infty > 0$
  \item $(V' \circ\phi)(t) \to 0$
  \item $\dot\phi(t),\,\ddot\phi(t) \to 0$
 \end{enumerate}
\end{prop}

\begin{proof}
 (i) By Lemma \ref{bounds} $H$ is monotonically decreasing and bounded from
 below by $H_0>0$, so convergence follows immediately. (ii) Using the decay
 of the matter terms and of $\dot\phi$ from Proposition \ref{mtdecay} as
 well as the convergence of $H$ just obtained, the Hamiltonian constraint
 \eqref{hamcon} yields $2(V\circ\phi)(t) \to n(n-1) H^2_\infty$ $(t
 \to \infty)$. (iii) Let $\phi_-:=\liminf_{t\to\infty} \phi(t)$ and
 $\phi_+:=\limsup_{t\to\infty} \phi(t)$ be the limes inferior and limes
 superior of $\phi$ in $\setrq$ respectively. If $\phi_-=\phi_+$, then
 $\phi(t)$ converges for $t\to\infty$ and the limit lies in the closure
 of $J$ in $\setrq$, so $(V' \circ\phi)(t)$ converges in $\setrq$ by
 assumption (P3).  Thus, suppose $\phi_- < \phi_+$ and choose $\phi_0$
 in the open interval $\ioo{\phi_-,\phi_+}$. Then, there exists a
 sequence $(s_n)$ in $I$ with $s_n\to\infty$ and $\phi(s_n)\to\phi_0$
 for $n\to\infty$ because $\phi$ is continuous. But then (ii) implies
 \[V_\infty = \lim_{t\to\infty} (V\circ\phi)(t) = \lim_{n\to\infty}
 V(\phi(s_n)) = V(\phi_0),\] so $V$ is constant on $\ioo{\phi_-,\phi_+}$
 and thus $V'$ vanishes on $\ioo{\phi_-,\phi_+}$. This means that for
 every neighborhood $N$ of zero, $U:=(V')^{-1}(N)$ is a neighborhood
 of the closure of ${\ioo{\phi_-,\phi_+}}$ in $\setr$. By construction,
 it follows that $\phi(t)\in U$ for $t$ sufficiently large and therefore
 $(V' \circ\phi)(t) \in N$ eventually. This shows the convergence of $(V'
 \circ\phi)(t)$ in $\setrq$ for $t\to\infty$ in any case. Employing the
 scalar field equation \eqref{sfeq} and using Proposition \ref{mtdecay}
 it follows that $-\ddot\phi$ converges to the same limit, so boundedness
 of $\dot\phi$ requires that limit to vanish.
\end{proof}

The results of this section show in particular that if the potential contains
a cosmological constant, \emph{i.e.} is bounded below away from zero, the
deceleration parameter $q:=-1-\dot H/H^2$ approaches $-1$ at late times,
so the expansion of the universe is accelerated exponentially. Moreover, the
density of ordinary matter $C(\phi)\rho$ as well as any anisotropy
$|\sigma|_\gamma /H$ vanish exponentially fast.

%
%

\section{Intermediate acceleration} \label{sec:int}

In the case in which the potential energy density of the scalar field
$V(\phi)$ can become arbitrarily small, $\inf\,(V \circ \phi)(I)=V_0=0$,
the dynamics is more subtle. Nevertheless, for potentials falling off
not too steeply, it is possible to prove late-time acceleration as well
as decay estimates for the curvature and matter terms. In contrast to the
findings in the case of positive $V_0>0$, the acceleration is in general
no longer exponential but may be asymptotically power-law. This will be
referred to as \emph{intermediate acceleration}. Of course, the solution
again is supposed to exist globally in time.

In general, \emph{i.e.} without having a positive lower bound $V_0$ on the
field's potential energy density, Proposition \ref{mtdecay} is weakened
to the following Proposition \ref{mtdecay*}.

\begin{prop} \label{mtdecay*}
 The spatial curvature $R_\gamma$, the shear $|\sigma|^2_\gamma$,
 the matter terms $C(\phi)\rho$, $C(\phi)|j|_\gamma$
 and $C(\phi) \tr\gamma S$ as well as the coupling term
 $c(\phi) \tr g T$ decay at least as $t^{-2}$, \emph{i.e.}
 \[R_\gamma,\,|\sigma|^2_\gamma,\,C(\phi)\rho,\,C(\phi)|j|_\gamma,\,C(\phi)
 \tr\gamma S,\,c(\phi) \tr g T = \mathcal{O}\left(t^{-2}\right) \quad
 (t \to \infty).\]

\end{prop}
 
\begin{proof}
 Using the same quantity $Z=n(n-1)H^2-\dot\phi^2-2V(\phi)$ as in the
 proof of Proposition \ref{mtdecay} it follows by Lemma \ref{bounds} that
 $\dot\phi(t) \to 0$ as $t \to \infty$ and so both $Z \geq 0$ and $\dot Z
 \leq -HZ$ eventually are valid. The Hamiltonian constraint \eqref{hamcon}
 in turn gives $Z \leq n(n-1)H^2$, so \[ \dot Z \leq -\frac1{\sqrt{n(n-1)}}
 Z^\frac32 \quad {\rm eventually} \] holds. Integrating this yields \[
 Z(t) = \mathcal{O} \left(t^{-2} \right) \qquad (t\to\infty), \] which
 establishes the claimed decay by (DEC) and (C).
\end{proof}

Since $V(\phi)$ is positive and continuous on $I$, the condition $V_0=0$
is equivalent to $\liminf_{t\to\infty} (V\circ\phi)(t) = 0$. Monotonicity of
$H$ and the constraint equation \eqref{hamcon} then give $\lim_{t\to\infty}
H(t)=0$ and, together with the Proposition \ref{mtdecay*} just proven,
that actually $\lim_{t\to\infty} (V\circ\phi)(t) = 0$. By presupposition
(P1) it follows that $J$ is unbounded and either $\lim_{t\to\infty}
\phi(t)=-\infty$ or $\lim_{t\to\infty} \phi(t)=\infty$. Without loss of
generality the latter will be assumed for the rest of this section.

Restricting the asymptotic steepness of the potential \[ \alpha :=
\limsup_{x\to\infty} \frac{|V'|}V (x) \] will allow more detailed asymptotics
to be obtained and late-time accelerated expansion to be shown. For this,
consider the quantity \[ W:= \frac{n(n-1) H^2}{2 V(\phi)}. \] According
to the equations \eqref{hamcon} and \eqref{hevl} it fulfills $W \geq 1$,
\begin{equation} \label{phidot}
 \frac{\dot\phi^2}{H^2} = n(n-1) \left( 1-\frac1W \right) -
 \frac{|\sigma|^2_\gamma}{H^2} + \frac{R_\gamma}{H^2} - \frac{2
 C(\phi)\rho}{H^2}
\end{equation} 
and its time derivative is given by
\begin{equation} \label{wdot}
 \begin{split}
  \dot W & = -\frac2{n-1} \Bigg[ \frac{\dot\phi^2}H + \frac12 (n-1)
  \frac{V'}V(\phi) \dot\phi \\
  & \quad + \frac{|\sigma|^2_\gamma}H -\frac1n \frac{R_\gamma}H +\frac1n
  \frac{C(\phi)}H (n\rho + \tr\gamma S) \Bigg] W
 \end{split}
\end{equation} 
respectively. Also note that with the elementary inequality $\sqrt{a-b} \geq
\sqrt a - b/\sqrt a$ for all $a>b\geq0$ the relation
\begin{multline} \label{phidotest}
  -\frac{\dot\phi^2}H - \frac{|\sigma|^2_\gamma}H + \frac1n \frac{R_\gamma}H
  - \frac1n \frac{C(\phi)}H (n\rho + \tr\gamma S) \leq -\sqrt{n(n-1)}
  \left( 1-\frac1W \right) |\dot\phi| \\
  {} - \left( 1-\frac1{\sqrt{n(n-1)}} \frac{|\dot\phi|}H \right)
  \frac{|\sigma|^2_\gamma}H + \left( \frac1n - \frac1{\sqrt{n(n-1)}}
  \frac{|\dot\phi|}H \right) \frac{R_\gamma}H \\
  {} - \left( \frac12 \frac{n-1}n -\frac1{\sqrt{n(n-1)}} \frac{|\dot\phi|}H
  \right) \frac{2C(\phi)}H
\end{multline} follows from \eqref{phidot}.

\begin{prop} \label{wlimsup}
 If $\alpha < 2/\sqrt{n(n-1)}$ then \[ \limsup_{t\to\infty} W(t) \leq \left(1
 - \frac12 \sqrt\frac{n-1}n \alpha \right)^{-1}.\]
\end{prop}

\begin{proof}
 Fix $\epsilon>0$ with $\alpha+2\epsilon<2/\sqrt{n(n-1)}$, then there is a
 $T\in I$ with $(|V'|/V)(\phi(t)) \leq \alpha+\epsilon$ for all $t\geq T$.
 Define \[ 1 < \delta := \left( 1- \sqrt\frac{n-1}n \frac{\alpha+2\epsilon}2
 \right)^{-1} < \frac n{n-1} \] and let $T \leq L \subset I$ be any
 subinterval of $I$ bounded below by $T$ on which $W$ is not less than
 $\delta$, $W|L\geq \delta$. For an arbitrary $t\in L$ distinguish two
 cases.  First, assume that $(|\dot\phi|/H)(t)\geq \sqrt{(n-1)/n}$, then
 \eqref{wdot} directly gives \[ \dot W(t) \leq \left( - \frac2{\sqrt{n(n-1)}}
 + \alpha +\epsilon \right) |\dot\phi(t)| W(t) \leq -\epsilon |\dot\phi(t)|
 W(t). \] Second, assume $(|\dot\phi|/H)(t)\leq \sqrt{(n-1)/n}$ instead,
 so \[ \frac1{\sqrt{n(n-1)}} \frac{|\dot\phi|}H \leq \frac1n \leq \frac12
 \frac{n-1}n \leq 1, \] then \eqref{phidotest} yields \[\dot W(t) \leq
 \left[ -2 \sqrt\frac n{n-1} \left( 1- \frac1\delta \right) + \alpha
 + \epsilon \right] |\dot\phi(t)| W(t) \leq -\epsilon |\dot\phi(t)|
 W(t).\] This, together, shows that on $L$ the inequality $\dot W \leq
 -\epsilon W |\dot\phi|$ holds, so $L$ is bounded because $\dot\phi$ is
 non-integrable. Since $W$ is in particular monotonically decreasing on
 any such $L$ the set where $W$ is bigger than $\delta$ is itself bounded,
 but this means that $\limsup_{t\to\infty} W(t) \leq \delta$.
\end{proof}

Combining the evolution and constraint equations \eqref{hamcon},\eqref{hevl}
with the relation \eqref{phidot} results in an upper bound on the
deceleration parameter $q$, namely
\begin{equation} \label{interq}
 q \leq n \left( 1-\frac1W \right) -1,
\end{equation}
leading to the following sufficient condition for accelerated expansion.
\begin{prop} \label{iaclcond}
 If $\alpha < 2/\sqrt{n(n-1)}$ then accelerated expansion occurs eventually.
\end{prop}

\begin{proof}
 From Proposition \ref{wlimsup} it is known that \[ \limsup_{t\to\infty}
 W(t) \leq \left(1 - \frac12 \sqrt\frac{n-1}n \alpha \right)^{-1} < \frac
 n{n-1},\] so $W < n/(n-1)$ eventually. By \eqref{interq} this means that
 $q<0$ eventually.
\end{proof}

Isotropization can be proven by requiring the potential to be ``flat'' at
infinity, which says that $\alpha = 0$.
\begin{prop} \label{isotrp}
 If the potential is flat at infinity, \emph{i.e.} $\lim_{x\to\infty}
 (V'/V)(x) = 0$, then the curvature and matter terms vanish faster than
 $H^2$, more precisely, as $t \to \infty$, \[\frac{\dot\phi}H(t),
 \frac{|\sigma|^2_\gamma}{H^2}(t), \frac{R_\gamma}{H^2}(t),
 \frac{C(\phi)\rho}{H^2}(t), \frac{C(\phi)|j|_\gamma}{H^2}(t),
 \frac{C(\phi)\tr\gamma S}{H^2}(t), \frac{c(\phi)\tr gT}{H^2}(t) \to 0.\]
\end{prop}

\begin{proof}
 When $\alpha=0$, Proposition \ref{wlimsup} implies $W(t) \to 1$ as $t \to
 \infty$ and equation \eqref{phidot} together with (DEC) and (C) makes the
 claims evident immediately.
\end{proof}

A question that arises in the treatment outlined so far is whether it
is possible to decide \emph{a priori}, \emph{i.e.} without referring
to the actual solution, to which of the two cases ($V_0>0$ or $V_0=0$)
the evolution belongs. As an answer to this question in general not only
depends on the form of the potential $V$ but also on the initial data, no
simple necessary and sufficient criterion can be expected. Instead, two
rather rough conditions leading to each of the cases will be presented,
that might be useful in some situations though. They can be employed to make
contact with the findings of \cite{Ren05}.

Take a barrier $B>0$ and consider a component $U$ of the set $\{x\in
J:V(x)\leq B\}$. By monotonicity of $H$ it is clear that if $H^2(t_0) \leq
2B/n(n-1)$ and $\phi$ belongs to $U$ initially, $\phi(t_0) \in U$, then $\phi$
stays in $U$, $\phi(I) \subset U$. Now the conditions can be stated as
\begin{lemma} \label{conpos}
 If $U$ is bounded then $V_0 > 0$.
\end{lemma}

\begin{proof}
 With $U$ its closure $\bar U$ in $\setr$ is bounded too and thus compact.
 Because of $V(\phi) \geq \bar V(\phi)$ and $(\bar V \circ \phi)(I) \subset
 \bar V(\bar U)$ it follows that $V_0 = \inf (V\circ\phi)(I) \geq \inf (\bar
 V\circ\phi)(I) \geq \inf \bar V(\bar U) >0$ from (P1).
\end{proof}

\begin{lemma} \label{connul}
 If $V'<0$ on $U$ and $U$ is bounded below within $J$, then
 $\lim_{t\to\infty} \phi(t) = J_+$. In particular, if the potential
 has negative derivative everywhere and $V(x)$ decays as $x\to \infty$
 then $V_0=0$.
\end{lemma}

\begin{proof}
 Let $U_\pm$ denote the endpoints of $U$ in $\setrq$. By presupposition,
 $U_-$ is an element of $J$ and thus of $U$ itself. It will now be
 shown that actually $U_+=J_+$. Assume that this is not the case, so
 $U_+<J_+$, then $U_+$ lies in the interior of $J$ and $V'<0$ on $U$
 implies that $U$ can be extended in $J$ beyond $U_+$ while remaining a
 connected subset of $\{x\in J:V(x)\leq B\}$. But this is a contradiction
 to $U$ being a component of that set. So indeed $U_+=J_+$. If $V_0>0$
 Proposition \ref{flddecay} gives $(V'\circ\phi)(t)\to 0$ for $t\to\infty$
 and therefore $\phi(t)\to J_+$ $(t\to\infty)$. If $V_0=0$ nothing is left
 to be shown. The additional statement is immediate.
\end{proof}

%
%

\section{Curvature-assisted acceleration} \label{sec:caa}

The results of the previous sections \ref{sec:exp} and \ref{sec:int} shall
now be applied exemplarily to the case of a scalar field with an explicit
coupling to the scalar curvature of space-time to demonstrate the mechanism
of curvature-assisted acceleration. For simplicity an exponential potential
is assumed although the method is not restricted to that case by any
means. By a conformal transformation the direct coupling of the field to the
Ricci curvature is first moved to the energy-momentum tensor where results
are obtained by virtue of the propositions proved so far. Transforming
back to the Jordan frame will then yield the desired statements.

On an interval $\jrd I:=\ico{\jrd{t}_0,\infty}$ suppose to have functions
$\jrd\phi \in C^2(\jrd I)$, $\jrd\rho \in C^1(\jrd I)$, $\jrd j \in
C^1(I,T^0_1\g)$ and families $\jrd\gamma \in C^2(\jrd I,T^0_2\g)$ and $\jrd
S \in C^1(\jrd I,T^0_2\g)$ of Riemannian metrics and symmetric tensors on
the Lie algebra $\g$ respectively such that $\jrd T:=\jrd\pi^\ast \jrd
S - \jrd\pi^\ast \jrd j \otimes d\jrd t - d\jrd t \otimes \jrd\pi^\ast
\jrd j + \jrd\rho\, d\jrd t \otimes d\jrd t$ satisfies the dominant and
strong energy conditions (DEC) and (SEC) while $\jrd g := \jrd\pi^\ast
\jrd\gamma - d\jrd t \otimes d\jrd t$ is a solution to the curvature-coupled
Einstein-scalar~field-matter (ESM) equations
\begin{align}
 \begin{split}
 (1-\xi \jrd\phi^2)\left( Rc_{\jrd g} - \frac12 R_{\jrd g} \jrd{g} \right)
  & = (1-2\xi) \del_{\jrd g} \jrd\phi \otimes \del_{\jrd g} \jrd\phi +
  \left( 2\xi - \frac12 \right) |\del_{\jrd g} \jrd\phi|^2_{\jrd g} \jrd
  g \\
  & \quad - 2\xi \jrd\phi \del^{\,2}_{\jrd g} \jrd\phi + 2\xi (\jrd\phi
  \Box_{\jrd g} \jrd\phi) \jrd g - \jrd V(\jrd\phi) \jrd g + \jrd T
  \label{cceseq}
 \end{split} \\
 \Box_{\jrd g} \jrd\phi - \xi R_{\jrd g} \jrd\phi - \jrd V'(\jrd\phi) & 
 = 0 \label{ccsfeq} \\
 \Div{\jrd g} \jrd T & = 0 \label{ccmteq}
\end{align}
on $\jrd M:=G \times \jrd I$ with potential $\jrd V$ of class $C^1$. Like
before, $\jrd\pi : \jrd M \to G$ and $\jrd t : \jrd M \to \jrd I$ are the
canonical projections, compositions with which are understood implicitly
when appropriate. The coupling constant $\xi$ is taken to be any non-zero
real number.

Assume $1-\xi \jrd\phi^2 >0$ and define a conformal factor \[\Omega :=
\sqrt[n-1]{1-\xi \jrd\phi^2} \in C^2(\jrd I).\] The transformation
functions
\begin{gather}
 p: \jrd I \to I,\quad \jrd t \mapsto \int_{\jrd{t}_0}^{\jrd t} \Omega \\
 P: \jrd J \to \setr,\quad \jrd x \mapsto \int_0^{\jrd x} \frac{\sqrt{1-\eta
 \xi x^2}}{1-\xi x^2} dx
\end{gather}
are smooth diffeomorphisms from $\jrd I$ onto $I:=p(\jrd I)$ and from \[ \jrd
J := \begin{cases}
 \ioo{-1/\sqrt\xi,1/\sqrt\xi} & \text{if $\xi$ is positive} \\
 \setr & \text{if $\xi$ is negative}
\end{cases} \] onto $\setr$ whose inverses will be denoted by $q$ and $Q$
respectively. For convenience the constant $\eta := 1-4\xi n/(n-1)$
is introduced. Note that the special case of conformal coupling is
characterized by the vanishing of $\eta$. Having
\begin{gather*}
 \gamma := (\Omega^2 \jrd\gamma) \circ q, \qquad \phi := P(\jrd\phi) \circ q \\
 \rho := (\Omega^{-2} \jrd\rho) \circ q, \qquad j := (\Omega^{-1} \jrd j)
  \circ q, \qquad S := \jrd S \circ q
\end{gather*}
the metric $g$, the field $\phi$ and the energy-momentum tensor $T$ built
up as described in section \ref{sec:kin} satisfy the matter-coupled ESM
equations \eqref{feseq}--\eqref{fmteq} on $M=G\times I$ with the smooth
coupling functions
\begin{align*}
 C & :=\frac1{1-\xi Q^2} & c & := \frac1{n-1}C'=\frac1{n-1} \frac{2\xi
 Q}{\sqrt{1-\eta \xi Q^2}} C
\end{align*}
and the transformed potential $V:= \jrd V(Q) C^\frac{n+1}{n-1}$. It is
easily seen that (SEC) and (DEC) hold for $T$ as well and the condition
(C) is fulfilled due to the boundedness of $2\xi Q/\sqrt{1-\eta \xi Q^2}$
on $\jrd J$.

Attention shall now be restricted to positive coupling constants $\xi >0$
and fields evolving in exponential potentials $\jrd V(x)=\lambda e^{-\kappa
x}$ for positive $\kappa,\lambda >0$ and all $x \in \setr$. Suppose $\jrd
H(\jrd t_0) > - (\ln \Omega)\dot{}(\jrd t_0)$ and define \[ \jrd\phi_\infty
:= \sqrt{\frac1\xi + \frac1{\kappa^2} \left( \frac{n+1}{n-1} \right)^2 } -
\frac1\kappa \frac{n+1}{n-1} \] so that $\phi_\infty := P(\jrd\phi_\infty)$
is the critical point of the transformed potential $V=\jrd V(Q)
C^\frac{n+1}{n-1}$. Then the following Proposition \ref{caa} holds true.
\begin{prop} \label{caa}
 In the limit $\jrd t \to \infty$, the field $\jrd\phi$ and its derivatives
 converge, \[ \jrd\phi(\jrd t) \to \jrd\phi_\infty,\quad \dot{\jrd\phi}(\jrd
 t) \to 0,\quad \ddot{\jrd\phi}(\jrd t) \to 0,\] so does the mean curvature
 $-\jrd H$, \[ \jrd H(\jrd t) \to \jrd H_\infty := \sqrt{\frac2{n(n-1)}
 \frac{\jrd V(\jrd\phi_\infty)}{1-\xi \jrd\phi_\infty^2}} \] and the
 curvature and matter terms vanish exponentially, \[ R_{\jrd\gamma},\
 |\jrd\sigma|^2_{\jrd\gamma},\ \jrd\rho,\ |\jrd j|_{\jrd\gamma},\
 \tr{\jrd\gamma} \jrd S = \mathcal{O}\left(e^{-(2-\delta) \jrd H_\infty
 \jrd t} \right)\] for any $\delta > 0$. Moreover, asymptotically, the
 expansion is accelerated exponentially \[ \jrd q = -1 -\frac{\dot{\jrd
 H}}{\jrd H^2} \to -1.\]
\end{prop}

\begin{proof}
 (i) The presupposition $\jrd H(\jrd t_0) > - (\ln \Omega)\dot{}(\jrd
 t_0)$ is equivalent to $H(0)>0$ while $0=\min I$. Furthermore, the
 transformed potential $V$ clearly fulfills the requirements (P1),
 (P2) and (P3) and $V(x)$ goes to infinity when $|x|$ does. Hence, it
 possesses a positive lower bound and Lemma \ref{conpos} ensures the
 image $\phi(I)$ being relatively compact which implies that $\Omega$
 is bounded away form zero. This renders $\Omega \notin L^1(\jrd I)$
 non-integrable and so the transformed evolution exists globally too,
 $I=\ico{0,\infty}$. The results of section \ref{sec:exp} thus apply.  (ii)
 From Proposition \ref{flddecay} convergence of $\phi(t)$ to $\phi_\infty$
 and therefore of $\jrd\phi(\jrd t)$ to $\jrd\phi_\infty$ follows. The decay
 of $\dot\phi$ and $\ddot\phi$ give those for the derivatives of $\jrd\phi$
 as well as convergence of $\jrd H$ and $\jrd q$.  The exponential decay
 of the curvature and matter terms $|\jrd\sigma|^2_{\jrd\gamma} = \Omega^6
 (|\sigma|^2_\gamma \circ p)$, $R_{\jrd\gamma} = \Omega^2 (R_\gamma \circ
 p)$, $\jrd\rho = \Omega^{n+1} [C(\phi)\rho \circ p]$, $|\jrd j|_{\jrd\gamma} =
 \Omega^{n+1} [C(\phi)|j|_\gamma \circ p]$ and $\tr{\jrd\gamma}
 \jrd S = \Omega^{n+1} [C(\phi)(\tr\gamma S) \circ p]$ is then directly
 obtained from Proposition \ref{mtdecay}.
\end{proof}

This result shows that an arbitrarily small positive coupling constant $\xi$
establishes a dynamics similar to the presence of a cosmological constant
with value \[ \Lambda_{\rm dyn} = \frac{\jrd V(\jrd\phi_\infty)}{1-\xi
\jrd\phi_\infty^2} \] although the potential lacks a positive lower
bound. It causes exponential acceleration to occur asymptotically as well
as exponentially fast isotropization and decay of matter independent of
the steepness $\kappa$ of the potential.

In the proof given above the particular form of the potential $\jrd V$ is
used mainly to ensure the existence of exactly one critical point of $V$
and thus to obtain convergence of the field $\jrd\phi$ and the conformal
factor $\Omega$ at late times. For assumptions (P1), (P2) and (P3) to
hold it is sufficient for instance to require the potential only to allow
for a $C^1$-extension to the closure of $\jrd J$ in $\setr$. Invoking
Lemma \ref{conpos} it can still be concluded that the field $\jrd\phi$
cannot approach the boundary of $\jrd J$ and hence the presumption $1-\xi
\jrd\phi^2 >0$ does not restrict the dynamics even in this more general
situation once it is fulfilled initially.

%
%

\section{Conclusions} \label{sec:con}

In this paper spatially homogeneous solutions to the Einstein equations in
the presence of a non-minimally coupled scalar field and ordinary matter were
investigated. Two different kinds of direct couplings, both to the matter
and to the scalar curvature of space-time, were considered with respect to
late-time acceleration and isotropization. The results are summarized in
the following theorems that are immediate consequences of the Propositions
\ref{mtdecay}, \ref{flddecay} and \ref{iaclcond}, \ref{isotrp} respectively.

\begin{theorem} \label{thmmin}
 Consider a solution of Bianchi type I--VIII of the Einstein equations
 together with a non-linear scalar field evolving in a positive potential
 and coupled directly to ordinary matter satisfying the dominant and strong
 energy condition. Suppose the potential to possess a positive lower
 bound and the assumptions of section \ref{sec:exp} to be fulfilled. If
 the solution is expanding initially and exists globally in the future
 then exponential acceleration and isotropization occur asymptotically.
\end{theorem}

\begin{theorem} \label{thmexp}
 Consider the same situation as in Theorem \ref{thmmin} above but suppose
 instead of a positive lower bound that the the asymptotic steepness
 $\alpha$ of the potential is less than $\sqrt{2/3}$. If the solution is
 expanding initially and exists globally in the future then accelerated
 expansion takes place eventually. Moreover if the potential is actually
 flat at infinity, $\alpha=0$, the model isotropizes as well.
\end{theorem}

The two statements just given are \emph{cosmic no-hair} theorems for
coupled quintessence. Note that the particular value of $\sqrt{2/3}$ in
the condition on the asymptotic steepness of the potential appears too in
\cite{Kit92} and \cite{Ren06}.

In the case in which the scalar field is non-minimally coupled to the
space-time curvature a conformal transformation was performed to cast the
model into a coupled quintessence scenario where Theorem \ref{thmmin} applied.
As a consequence a late-time version of Tsujikawa's \emph{curvature-assisted}
mechanism \cite{Tsu00} was obtained.

\begin{theorem} \label{thmcaa}
 Consider a solution of Bianchi type I--VIII of the Einstein equations
 together with a non-linear scalar field coupled directly to the scalar
 curvature of space-time and evolving in an arbitrary exponential potential
 in the presence of ordinary matter satisfying the dominant and strong
 energy condition.  Suppose that for a positive coupling constant $\xi$
 the conditions $1-\xi\phi^2>0$ and $H>-[\ln(1-\xi\phi^2)]\dot{\,}/(n-1)$
 on the field $\phi$ and the expansion factor $H$ are fulfilled initially.
 If the solution exists globally in the future then $1-\xi\phi^2>0$ holds
 at any time and exponential acceleration and isotropization take place
 asymptotically.
\end{theorem}

This result might be physically interesting since it provides a positive
cosmological constant although the potential does not include one. Even more,
this is true independent of the steepness of the potential, \emph{i.e.}
the expansion will be accelerated exponentially in potentials that would not
even allow for power-law inflation in the minimally coupled case.

%
%

\end{document}